\documentclass[twocolumn,pra,amsmath,amssymb,superscriptaddress]{revtex4-1}
\usepackage[utf8]{inputenc}
\usepackage{amsfonts,amsmath,amssymb,mathtools}
\usepackage{graphicx}
\usepackage{bm} 
\usepackage{mathrsfs}
\usepackage{subfigure}
\usepackage{textcomp}
\usepackage{microtype}
\usepackage{hyperref}
\usepackage[flushleft]{threeparttable}
\usepackage[per-mode=symbol]{siunitx}
\usepackage[version=3]{mhchem}

\usepackage{booktabs}
\graphicspath{{figures/}}

\DeclareSIUnit\intensity{\watt\per\centi\meter\squared}
\DeclareSIUnit\fieldstrength{\volt\per\centi\meter}
\DeclareSIUnit\u{\atomicmassunit}
\usepackage{xspace}

\DeclareUnicodeCharacter{2009}{\,}

\newcommand{\singlet}{$1^1\Sigma_g^+$\xspace}
\newcommand{\triplet}{$1^3\Sigma_u^+$\xspace}

\newlength{\figwidth}
\newlength{\figwidthwide}
\let\orgautoref\autoref
\providecommand{\Autoref}{%
  \def\equationautorefname{Equation}%
  \def\figureautorefname{Figure}%
  \def\subfigureautorefname{Figure}%
  \def\tableautorefname{Table}
  \def\sectionautorefname{Section}%
  \orgautoref}
\renewcommand{\autoref}{%
  \def\equationautorefname{Eq.}%
  \def\figureautorefname{Fig.}%
  \def\subfigureautorefname{Fig.}%
  \def\sectionautorefname{Sec.}%
  \orgautoref}

\usepackage{color}
\usepackage[normalem]{ulem}
\definecolor{darkgreen}{rgb}{0.0,0.7,0.0}

\newcommand{\avg}[1]{\langle #1 \rangle}

\DeclareSIUnit{\mbar}{\milli\bar}
\DeclareSIUnit{\eV}{\electronvolt}
\DeclareSIUnit{\meV}{\milli\electronvolt}
\DeclareSIUnit\u{\text{u}}
\DeclareSIUnit\e{\text{e}}

\DeclareSIUnit\mJ{\milli\joule}
\DeclareSIUnit\uJ{\micro\joule}
\DeclareSIUnit\mW{\milli\Watt}
\DeclareSIUnit\mm{\milli\meter}
\DeclareSIUnit\um{\micro\meter}
\DeclareSIUnit\nm{\nano\meter}
\DeclareSIUnit\ns{\nano\second}
\DeclareSIUnit\ps{\pico\second}
\DeclareSIUnit\fs{\femto\second}

\begin{document}

\title{Femtosecond-and-atom-resolved solvation dynamics of a Na$^+$ ion in a helium nanodroplet} 

\author{Simon H. Albrechtsen}
\affiliation{Department of Chemistry, Aarhus University, Langelandsgade 140, DK-8000 Aarhus C, Denmark}

\author{Jeppe K. Christensen}
\affiliation{Department of Chemistry, Aarhus University, Langelandsgade 140, DK-8000 Aarhus C, Denmark}

\author{Christian E. Petersen}
\affiliation{Department of Physics and Astronomy, Aarhus University, Ny Munkegade 120, DK-8000 Aarhus C, Denmark}

\author{Constant A. Schouder}
\affiliation{Université Paris-Saclay, CNRS, Institut des Sciences Moléculaires d’Orsay, 91405 Orsay, France}

\author{Pedro Javier Carchi-Villalta}
\affiliation{Institute of Fundamental Physics, CSIC (IFF-CSIC), Serrano 123, 28006 Madrid, Spain}

\author{Iker S\'anchez-P\'erez}
\affiliation{Institute of Fundamental Physics, CSIC (IFF-CSIC), Serrano 123, 28006 Madrid, Spain}

\author{Massimiliano Bartolomei}
\affiliation{Institute of Fundamental Physics, CSIC (IFF-CSIC), Serrano 123, 28006 Madrid, Spain}

\author{Tom\'as Gonz\'alez-Lezana}
\affiliation{Institute of Fundamental Physics, CSIC (IFF-CSIC), Serrano 123, 28006 Madrid, Spain}

\author{Fernando Pirani}
\affiliation{Dipartimento di Chimica, Biologia e
Biotecnologie, Universit\'a di Perugia, 06123 Perugia, Italy}

\author{Henrik Stapelfeldt}
\email[]{henriks@chem.au.dk}
\affiliation{Department of Chemistry, Aarhus University, Langelandsgade 140, DK-8000 Aarhus C, Denmark}

\date{\today}


\begin{abstract}

Recently, it was shown how the primary steps of solvation of a single Na$^+$ ion, instantly created at the surface of a nanometer-sized droplet of liquid helium, can be followed at the atomic level [Albrechtsen \textit{et al.}, Nature $\textbf{623}$, 319 (2023)]. This involved measuring, with femtosecond time resolution, the gradual attachment of individual He atoms to the Na$^+$ ion as well as the energy dissipated from the local region of the ion.
In the current work, we provide a more comprehensive and detailed description of the experimental findings of the solvation dynamics, and present an improved Poisson-statistical analysis of the time-resolved yields of the solvation complexes, Na$^+$He$_n$. For droplets containing an average of 5200 He atoms, this analysis gives a binding rate of $1.84\pm0.09$ atoms/ps for the binding of the first five He atoms to the Na$^+$ ion. Also, thanks to accurate theoretical values for the evaporation energies of the Na$^+$He$_n$ complexes, obtained by Path Integral Monte Carlo methods using a new potential energy surface presented here for the first time, we improve the determination of the time-dependent removal of the solvation energy from the region around the sodium ion. We find that it follows Newton’s law of cooling for the first 5 ps. Measurements were carried out for three different average droplet sizes, $\avg{N_D} = $ 9000, 5200 and 3600 helium atoms, and differences between these results are discussed.

\end{abstract}

\maketitle 

\section{Introduction}\label{sec:intro}

Solvation of ions is a fundamental and omnipresent process in many chemical and biological systems~\cite{marcus_ions_2015}. For instance, the detailed structure of the solvation shell of cations in water is believed to play an important role for how e.g. \ce{K+} ions pass through channels in cell membranes~\cite{marcus_effect_2009} and hydrated \ce{Na+} ions can influence the structure and arrangement of macromolecules like DNA~\cite{feig_sodium_1999}. Ions can also be solvated in liquid helium. This liquid may not be the first thing that comes to one’s mind as a solvent, but it is actually rather efficient for solvating cations~\cite{anderlanSolvationNaTheir2012}. As an example, the energy of a single isolated \ce{Na+} ion is lowered by 0.40 eV when embedded in a liquid helium nanodroplet containing 2000 He atoms~\cite{garcia-alfonso_time-resolved_2024}. Although this is considerably less than the energy an isolated \ce{Na+} ion is lowered by when dissolved in room temperature water ($\sim$~4.0~eV, the solvation energy)~\cite{marcus_ions_2015}, it is still sufficiently large that \ce{Na+} ions are well stabilized in liquid helium. The same is true for other alkali ions as well as for many other ionic species~\cite{glaberson_impurity_1975,tabbert_optical_1997,gonzalez-lezanaSolvationIonsHelium2020}.

Solvation of ions in liquid helium has been the subject of a large number of studies. Early reports, starting with the theoretical work of Atkins in 1959~\cite{atkins_ions_1959} and experimental work from various groups~\cite{williams_ionic_1957,reif_study_1960,johnson_positive_1972} addressed samples of bulk liquid helium and were motivated by the prospect of using ion mobility measurements as a microscopic probe of the special properties of the superfluid liquid.
The introduction of helium nanodroplets doped with atoms, molecules and clusters in the 1990s~\cite{toenniesSuperfluidHeliumDroplets2004,choiInfraredSpectroscopyHelium2006,stienkemeierSpectroscopyDynamicsHelium2006}, strongly expanded the scope of studies of ions in liquid helium~\cite{mauracherColdPhysicsChemistry2018} including infrared spectroscopy of a variety of cation species~\cite{florezIRSpectroscopyProtonated2015,davies_infrared_2019,singh_infrared_2023}, access to large biomolecular ions such as small proteins~\cite{bierauCatchingProteinsLiquid2010}, characterization of the helium solvation shells around e.g. alkali and alkaline earth cations through combinations of mass spectrometry and high-level theoretical modelling~\cite{doppnerIonInducedSnowballs2007,RLAOPBHCGHBSG:PCCP18,gonzalez-lezanaSolvationIonsHelium2020,zunzunegui-bru_observation_2023}, rotation of molecular cations~\cite{daviesOnsetRotationalDecoupling2023}, vibrational wave packet spectroscopy~\cite{qiang_femtosecond_2024}, ion-molecule reactions~\cite{farnik_ion-molecule_2004,denifl_ionmolecule_2009,renzler_observation_2016} and solvation dynamics of cations~\cite{lealPicosecondSolvationDynamics2014,albrechtsenObservingPrimarySteps2023,calvo_concurrent_2024,garcia-alfonso_time-resolved_2024,garcia-alfonso_revisiting_2024}. The latter is the subject of this paper and we will discuss femtosecond time-resolved solvation dynamics of a single \ce{Na+} ion in a helium droplet, atom-by-atom.

As described in the next section, our experimental method relies on starting the solvation by ionizing a Na atom, residing on the surface of a helium droplet, by a femtosecond laser pulse. Therefore, we find it important to mention early work addressing ionization of helium-surface-located atoms and the subsequent submersion of the atomic cation. In 2010, Ernst and coworkers experimentally showed that a Rb atom, initially bound at the surface of a helium droplet is immersed into the droplet, following two-photon ionization with a nanosecond laser pulse~\cite{theisenFormingRbSnowballs2010}. Similar conclusions were reached by Drabbels and coworkers in measurements of \ce{Ba+} ions created by 1-photon ionization of Ba atoms located at the droplet surface~\cite{zhangCommunicationBariumIons2012}. A few years later, time-dependent density functional theory (DFT) simulations provided picosecond time-resolved insight into how a \ce{Rb+} ion moves from the surface of a droplet to the interior while it develops solvation shells of helium atoms~\cite{lealPicosecondSolvationDynamics2014}. Other papers also treated laser-induced ionization of neutral precursor atoms as a means of initiating solvation of cations in helium droplets~\cite{theisen_cs_2011,lealDynamicsPhotoexcitedBa2016} but none of these works suggested a strategy for experimentally exploring the solvation dynamics in real time.

\begin{figure*}
    \includegraphics[width = 17.8 cm]{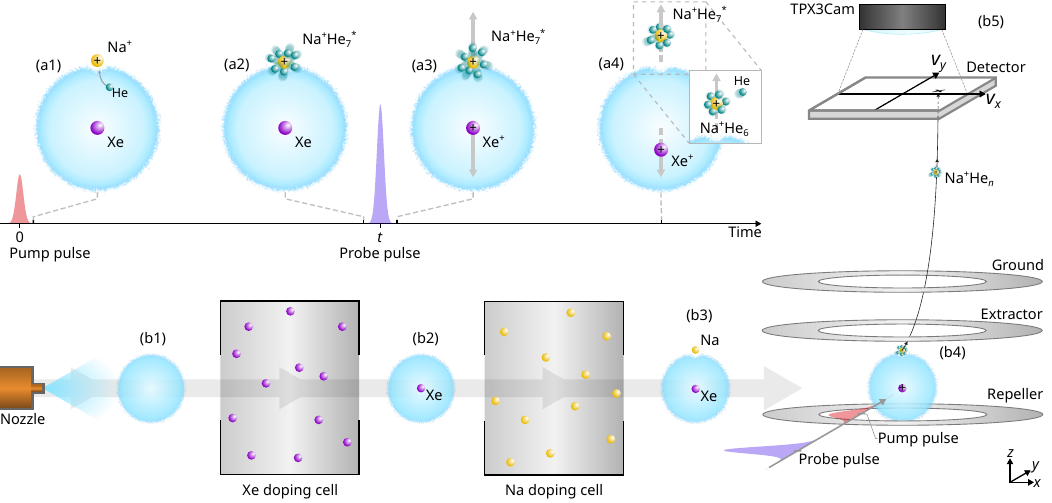}
    \caption{(a1)-(a4): Principle of the experiment. (a1): The initial ionization of the Na atom by the pump pulse. (a2): The formation of a \ce{Na+He_{$N$}} solvation complex. (a3)-(a4): The probe process in which the Xe atom is ionized and the \ce{Na+He_{$N$}} complex is repelled from the droplet. Panel (a4) illustrates that this complex may lose He atoms if its internal energy is sufficiently high. (b1)-(b5): Schematic of the experimental setup. (b1)-(b3): The part where the He droplets are formed and doped with a Na and a Xe atom. (b4)-(b5): The part where the interaction with the laser pulses happens and the \ce{Na+He_{$n$}} ions are projected onto the detector by the velocity map imaging spectrometer.
 \label{fig:expPrinciple}}
\end{figure*}

\section{Principle of the experiment}\label{sec:principle}

The starting point for the measurement is nanodroplets of liquid helium each containing one Na atom and one Xe atom. The heliophobic Na is weakly bound at the surface while the heliophilic Xe is located in the interior~\cite{poms_helium_2012}, see illustration in \autoref{fig:expPrinciple}(b3). Now the doped He droplets are irradiated by a 65 fs pump laser pulse, \autoref{fig:expPrinciple}(a1) and (b4)), which is strong enough to ionize the Na atom through multiphoton absorption, yet weak enough to avoid ionization of the Xe atom because its ionization potential, 12.1 eV, is significantly higher than that of Na, 5.1 eV. In this way, the \ce{Na+} ion is created at a well-defined time that defines the beginning of the solvation process of the cation, see \autoref{fig:expPrinciple}(a1). In addition, the solvation process also starts at a well-defined ion-solvent geometry because the distance from the Na atom, and thus from the \ce{Na+} ion immediately after its creation, to the closest part of the droplet is $\sim$ 5.0 {\AA} as shown by simulations based on He density functional theory~\cite{albrechtsenObservingPrimarySteps2023,garcia-alfonso_time-resolved_2024}. This well-defined geometry is critical for our ability to record the solvation process with sub-picosecond time-resolution~\footnote{Previous studies of the solvation of excited molecules on Argon clusters found multiple solvation timescales due to different solvation sites~\cite{awali_time_2013} - a complication avoided in our experiment due to the uniformity of the droplet dimple that the \ce{Na} atom sits in.}.

Time-dependent DFT simulations~\cite{albrechtsenObservingPrimarySteps2023,garcia-alfonso_time-resolved_2024} and ring-polymer molecular dynamics~\cite{calvo_concurrent_2024} predict that the newly formed \ce{Na+} ion gradually binds to He atoms in the droplet and that the energy released thereby is gradually dissipated. Our experimental strategy for following the solvation process relies on single-ionization of the Xe atom by multiphoton absorption from a 90 fs probe pulse arriving at time $t$ after the pump pulse, see \autoref{fig:expPrinciple}(a2)-(a3). The \ce{Xe+} ion created in this manner immediately exerts a repulsive force on \ce{Na+}, due to the repulsive Coulomb interaction between the two cations. This force will push the \ce{Na+} ion along with the number, $N$, of He atoms that are sufficiently tightly bound to it at this time,
in \autoref{fig:expPrinciple} $N = 7$ as an example, away from the droplet surface. It is possible that this complex, \ce{Na+He_{$N$}}, sheds one or more helium atoms after leaving the droplet surface, see \autoref{fig:expPrinciple}(a4) and discussion below. The resulting \ce{Na+He_{$n$}} ions, $n \leq N$, are directed towards a detector where their arrival time and velocity component in the detector plane are recorded, see \autoref{fig:expPrinciple}(b4)-(b5) and description in \autoref{sec:setup}. With this detection scheme, we are able to record the yield of \ce{Na+He_{$n$}} for $n$ in the range 0--25. We note that $n$ denotes the number of He atoms attached to the \ce{Na+} ion as recorded by the detector, while $N$ denotes the number of He atoms attached to the ion on the surface of the He droplet. The measurements are repeated for a number of delays, $t$, between the pump and the probe pulse and thus we obtain the \ce{Na+He_{$n$}} ion yields as a function of time. As discussed below, these observables allow us to determine the time-dependence of both the number of He atoms bound to the \ce{Na+} ion and of the energy dissipated as the cation is solvated.

\section{Experimental Setup}\label{sec:setup}
\Autoref{fig:expPrinciple}(b1)-(b4) show a schematic of the experimental setup. Helium gas with a purity of 99.9999\% regulated to a pressure of \SI{50}{\bar} is let into a nozzle that is cooled by a cryogenic cold-head (Sumitomo heavy industries RDK-408D2), and let through a \SI{5}{\um}--diameter round orifice (Plano Platin-Iridium A0200P) into a vacuum chamber maintained at a pressure of $\sim\SI{1e-4}{\mbar}$ by two Turbo pumps (Pfeiffer ATH 2303M and Edwards STP-iXR2206). Hence, the helium gas expands, cools down further and condenses to create a continuous beam of superfluid helium droplets with an expected temperature of \SI{0.37}{\kelvin}\cite{toenniesSuperfluidHeliumDroplets2004,shepperson_strongly_2017,pickeringAlignmentMathrmCS2019}, see \autoref{fig:expPrinciple}(b1). The size distribution of the droplets is expected to be a log-normal distribution~\cite{harms_density_1998}, and its mean value is determined by the nozzle temperature, regulated with a PID controller (Lakeshore Model 331) in the range \SIrange{10}{25}{\kelvin}~\cite{toenniesSuperfluidHeliumDroplets2004}. The three nozzle temperatures used in the experiment (with the corresponding average number of helium atoms and average radii of the droplets) were, \SI{20}{\K} ($\avg{N_D} = 3600$, $\avg{R} = \SI{3.3}{\nm}$), \SI{18}{\K} ($\avg{N_D} = 5200$, $\avg{R} = \SI{3.7}{\nm}$) and \SI{16}{\K} ($\avg{N_D} = 9000$, $\avg{R} = \SI{4.5}{\nm}$).

The droplet beam goes through a \SI{2}{\mm} skimmer into the doping vacuum chamber, pumped by a turbomolecular pump (Edwards STP-iX455). Here, the droplets pass through two doping cells where they preferentially pick up a single Xe atom and a single Na atom (see \autoref{fig:expPrinciple}(b2)-(b3)). Xenon gas is leaked into the first cell using a needle valve (Kurt J. Lesker LVM Series) while an appropriate amount of Na gas is obtained by heating a lump of sodium metal placed in the second cell to \SI{180}{\degreeCelsius} using a resistive heating coil and a PID controller (Eurotherm 3216). The doping levels of the droplets are set by controlling the pressure of gas in these two cells, and lead to an approximately Poissonian distribution of the number of dopants~\cite{toenniesSuperfluidHeliumDroplets2004}. Consequently, some of the droplets will be doped with Xe dimers or with Na dimers rather than with one Xe atom and one Na atom. Below we discuss how to discriminate against signal created by ionization of the homodimers. The doped droplets fly through a second \SI{2}{\mm} skimmer into the target chamber, maintained at a pressure of \SI{1e-9}{\mbar} by a turbopump (Edwards STP-iXR1606). Here, they enter the center of a velocity map imaging (VMI) spectrometer~\cite{chandlerTwoDimensionalImaging1987,eppinkVelocityMapImaging1997}, where they are crossed by two focused laser beams (see \autoref{fig:expPrinciple}(b4)).

The femtosecond laser system used is a Spectra Physics Spitfire Ace (central wavelength \SI{800}{\nm}, pulse duration \SI{50}{\fs}, pulse energy \SI{5}{\mJ}, repetition rate \SI{1}{\kilo\hertz}). The laser beam is split in two to create the pump and probe beams. The pump beam consists of part of the fundamental output from the laser, while the probe beam is created by second harmonic generation in a BBO crystal (\SI{0.5}{\mm} thickness) to yield pulses centered at \SI{400}{\nm}. A linear translation stage (Newport UTM100PP.1) is used to control the time delay, $t$, between the pump and probe pulses. The laser beams, linearly polarized along the x-direction (\autoref{fig:expPrinciple}(b4)), are recombined to be collinear using a dichroic mirror, and focused in the center of the VMI spectrometer with an achromatic doublet lens (focal length \SI{300}{\mm}). The focused spot sizes, $w_0$, are measured with a beam profiler (FemtoEasy LP6.3 UV), the pulse energies, $E_\text{pulse}$, with a powermeter (ThorLabs S310C), and the pulse durations, $\tau$, by cross correlation. The peak intensity of the pump pulse is \SI{5.8e13}{\W\per\cm^2} ($w_0$ = \SI{24}{\um}, $E_\text{pulse}$ = \SI{35}{\uJ}, $\tau$ = \SI{65}{\fs}), and \SI{9.1e13}{\W\per\cm^2} for the probe pulse ($w_0$ = \SI{16}{\um}, $E_\text{pulse}$ = \SI{35}{\uJ}, $\tau$ = \SI{90}{\fs}).

The VMI spectrometer projects the ions created by the laser pulses onto the detector consisting of a pair of microchannel plates backed by a P47 phosphor screen. The voltages on the repeller (\SI{6000}{\V}), extractor (\SI{3830}{\V}) and ground (\SI{0}{\V}) electrodes in the spectrometer are set to achieve velocity focusing. When the ions hit the MCP detector, they release a cascade of electrons, which in turn hit the phosphor screen causing it to emit light from a spot corresponding to the ion hit. The phosphor screen is imaged by a TPX3Cam~\cite{nomerotski_imaging_2019,fisher-levine_timepixcam_2016,zhao_coincidence_2017} event-based detector, which records the position and the time-of-arrival (ToA) and time-over-threshold (ToT) of each detector pixel that receives light intensity over a certain threshold. Each such pixel event is corrected for time-walk effects~\cite{pitters_time_2019,brombergerShotbyshot250KHz2022}, and hereafter clustered and centroided using an optimized implementation of the DBSCAN algorithm~\footnote{The custom DBSCAN implementation is publicly available at: \url{https://github.com/laulonskov98/pixel_centroiding_DBSCAN}}. In this way, the time-of-flight (ToF), and spatial coordinates, $x$ and $y$, of each ion hit are determined. The ToF is calibrated to give the $m/q$ value of each ion hit, and the $x$ and $y$ coordinates are calibrated to give the velocities in the detector plane, $v_x$ and $v_y$, using the well-known velocities of \ce{Na+} recoil ions from laser-induced Coulomb explosion of the \singlet and \triplet states of the \ce{Na2} dimer~\cite{kristensenQuantumStateSensitiveDetectionAlkali2022,kristensen_laser-induced_2023}.

The preprocessed data are then a list with the $v_x$, $v_y$, $m/q$ values and a laser shot index for each ion hit. Ion hits are recorded over a set period of time for a range of pump-probe delays between \SI{0.2} and \SI{20}{\ps}, which constitutes one run of the experiment. We can represent the ion hits for a given $m/q$ value as a function of $v_x$ and $v_y$. These 2-dimensional plots are referred to as 2D velocity images.
As mentioned above, three runs were done for three different droplet sizes ($\avg{N_D} =$ 3600, 5200, and 9000).

\section{Results and discussion}\label{sec:results}

\subsection{Time-dependent ion yields}
\begin{figure}
    \includegraphics[width = 8.6 cm]{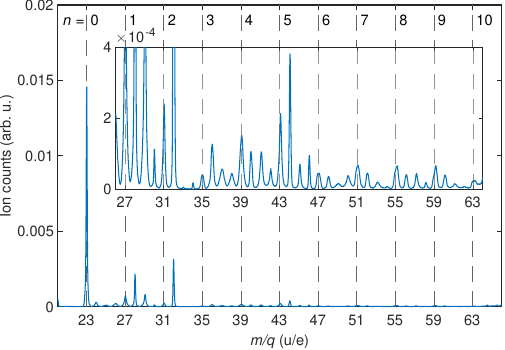}
    \caption{Mass-over-charge ($m/q$) spectrum integrated over all pump-probe delays for the ions recorded with droplet size $\avg{N_D} = 5200$. Only the range of $m/q$ from 23 to 63 is displayed. The droplets were doped with both \ce{Na} and \ce{Xe} atoms. The $m/q$ ratios corresponding to different \ce{Na+He_{$n$}} complexes are marked by the dashed grey lines. The inset shows a magnified view of data with $m/q$ between 27 and 63. \label{fig:massSpectrum}}
\end{figure}

\Autoref{fig:massSpectrum} shows a mass spectrum, obtained from the ToF of all ion hits detected, for the range \SI{20}{\u\per\e} $< m/q <$ \SI{66}{\u\per\e}. The average number of He atoms in the droplets was 5200. The most intense peak, $m/q$ = \SI{23}{\u\per\e}, is assigned as bare \ce{Na+} ions. At larger $m/q$ values, we observe a series of peaks centered at $m/q$ = $(23 + 4n)\si{\u\per\e}$. These $m/q$ ratios correspond to those of the \ce{Na+He_{$n$}} complexes, but there could potentially also be contributions to the peaks from e.g. hydrocarbon ions formed by ionization of small amounts of residual gas in the target chamber or from droplets doped with impurities.

\begin{figure}
    \includegraphics[width = 8.6 cm]{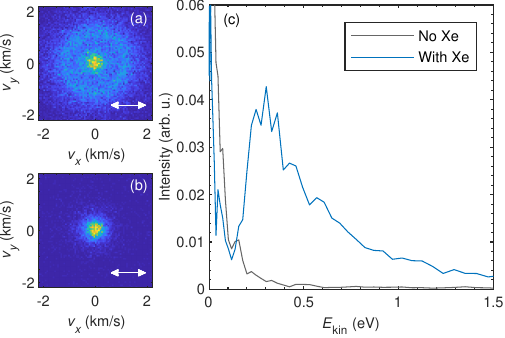}
    \caption{2D velocity images of $m/q = \SI{35}{\u\per\e}$ ions (corresponding to \ce{Na+He_3}) obtained from droplets with $\avg{N_D} = 5200$, doped with (a) both \ce{Na} atoms and \ce{Xe} atoms and (b) with \ce{Na} atoms only. The data shown are the sum of all time steps recorded. The corresponding kinetic energy distributions are shown in (c). The white arrows in (a) and (b) indicate the polarization axis of the laser pulses. \label{fig:image}}
\end{figure}

To get more insight, we analyze the 2D velocity images, recorded for helium droplets doped with both a Na and a Xe atom and for droplets doped with only a Na atom. As an example, \autoref{fig:image}(a) and (b) display the velocity image of the $m/q$ = $(23 + 4\cdot 3)~\si{\u\per\e}$ ions for these two droplet doping conditions. Figure \ref{fig:image}(a) exhibits a broad ring in the outermost part of the image, a structure that is absent in \autoref{fig:image}(b). The ring manifests itself as a broad peak with a center position of $\sim\SI{0.30}{\eV}$ in the kinetic energy distribution, blue curve in \autoref{fig:image}(c), obtained from Abel inversion of the 2D image, using the MEVIR algorithm~\cite{dick_inverting_2014}, and Jacobian transformation of the resulting velocity distribution as described in Ref.~\cite{kristensenQuantumStateSensitiveDetectionAlkali2022}. In comparison, we note that the average radius of the helium droplets is \SI{3.7}{\nm}. Thus, if a \ce{Na+He_3} is ejected from the droplet surface due to electrostatic repulsion from the \ce{Xe+} ion, assumed to be located in the droplet center, it acquires a final kinetic energy of $\sim\SI{0.31}{\eV}$~\footnote{We assume momentum conservation $m\left(\ce{Na+He_3}\right) \cdot v\left(\ce{Na+He_3}\right) = m\left(\ce{Xe+}\right) \cdot v\left(\ce{Xe+}\right)$ with $m\left(\ce{Xe+}\right) = \SI{130}{\u}$}. The radius of the droplets is expected to follow a log-normal size distribution~\cite{harms_density_1998}, with 95\% lying in the range \SIrange{2.6}{5.2}{\nm}, which will give the \ce{Na+He_3} complex a kinetic energy in the range \SIrange{0.22}{0.44}{\eV}. Deviations of the location of the \ce{Xe} atom from the droplet center, the resolution of the VMI spectrometer, and repulsions from a minor fraction of the \ce{Xe} atoms being doubly ionized, causes the experimental kinetic energy distribution to be broader than this. In the end, we assign the ions with energies in the range \SIrange{0.15}{1.6}{\eV} as those that were ejected from the droplet due to Coulomb repulsion from a \ce{Xe} ion, i.e. the ions coming from the process described in \autoref{sec:intro}.

\begin{figure*}
    \includegraphics[width = 17.8 cm]{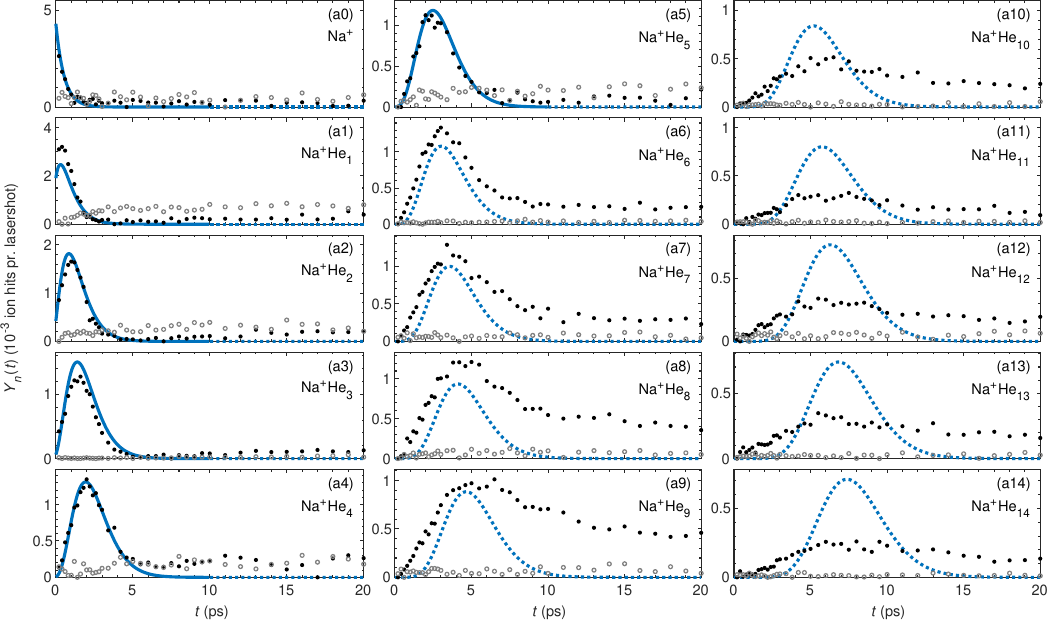}
    \caption{Black dots: Yields of different \ce{Na+He_{$n$}} ions, $Y_n$($t$),  measured as a function of pump-probe delay, $t$, using droplets doped with both a Na and a Xe atom and with $\avg{N_D} = $ 5200. Gray circles:  The same ion yields but recorded for droplets doped only with \ce{Na} atoms. Full blue lines: Results from the fit to the Poisson model. The dotted blue lines show the extrapolation of this fit beyond the fitted range, see text for further details.  \label{fig:yields}}
\end{figure*}

For each ion species with $m/q = (23 + n\cdot 4)\si{\u\per\e}$, $n=0-25$, we analyze the radial velocity distribution from the Abel inverted 2D velocity image and assign the radial range that corresponds to the \ce{Na+He_$n$ - Xe+} Coulomb repulsion channel. We cannot measure complexes larger than \ce{Na+He_{25}}, as these overlap with the isotopes of \ce{Xe+} and \ce{Xe+He_{$i$}} in the $m/q$ spectrum. By recording 2D velocity images of the ions for numerous delays between the pump and the probe pulse, we determine the \ce{Na+He_{$n$}} ion signal as a function of time, denoted $Y_n$($t$). The result is shown as filled black dots in \autoref{fig:yields} for $n = 0-13$. We note that for each curve, a small constant signal has been subtracted to remove a minor contribution from background ions that is still present despite the radial velocity selection of the ions. To justify this procedure, we repeated the measurement for droplets doped with \ce{Na} atoms only. Thereby, the \ce{Na+He_$n$ - Xe+} repulsion channel is removed and the ions detected should reveal the residual background contribution. The results, displayed as gray circles in \autoref{fig:yields}, show that without \ce{Xe} in the droplets the ion yields are small and, importantly, have no time dynamics. The exception is the \ce{Na+He_1} and \ce{Na+He_2} signals, \autoref{fig:yields}(a1-a2), which rise slightly during the first \SI{5}{\ps}. We believe this is due to pump-pulse-induced dynamics in the \ce{Na2} dimers, leading to \ce{Na+} ions that can bind to one or two He atoms before leaving the droplet. In practice, these background signals are essentially negligible and do not affect the $Y_n$($t$) signals in any significant manner.

The resulting time-dependent \ce{Na+He_{$n$}} ion yields, $Y_n$($t$), constitute the basic experimental observables. \Autoref{fig:yields} shows that for $n$~$>$~0, $Y_n$($t$) reaches a maximum and then decreases monotonically. The position of the maximum steadily shifts to larger times for increasing values of $n$, an observation showing that helium atoms gradually attaches to the \ce{Na+} ion. To quantify the attachment dynamics, we need a model for comparison with the experimental data. To apply and interpret the model, introduced in \autoref{sec:results:attachRate}, it is important to understand how the evaporation energy for the \ce{Na+He_{$N$}} complexes varies as a function of the number of He atoms, $N$. Therefore, the next section presents our calculation of the evaporation energies using a specially designed potential energy surface (PES).

\subsection{Evaporation energy of \ce{Na+He_{$N$}}}\label{sec:EvaporationEnergy}
In order to properly estimate the structural and energetic features of the \ce{Na+He_{$N$}} complexes, an accurate PES has been built.
The obtained force field is based on the sum of two-body (2B) and three-body (3B) non-covalent interaction contributions.
For the 2B contribution describing the He--He interaction we have employed
the corresponding expression from Ref.~\cite{AS:JCP91},
 while that for the \ce{Na+}--He interaction coupled cluster with single, double and perturbative triple excitations
[CCSD(T)] calculations were carried out by using the Molpro2012.1 package~\cite{Molpro}.
In particular, the interaction energies related to the  ground singlet state of He approaching the Na$^{+}$ ion  have been estimated at the CCSD(T) level in the complete basis set (CBS) limit.
In such computations  counterpoise corrected~\cite{Boys:70} interaction energies obtained with the aug-cc-pVQZ and aug-cc-pV5Z~\cite{Dunning} basis sets have been properly combined~\cite{Halkier:98,Halkier:99} to obtain  reliable extrapolated CBS values.
Once the above described {\it ab initio} results are obtained,
the \ce{Na+}--He interaction is analytically represented by means of
the improved Lennard Jones (ILJ) formulation given by~\cite{PBRCCV:PCCP08}:

\begin{equation}
\label{ILJ}
V(r) = \epsilon
\left[
{\frac{m}{n(r) \! - \!m}} \left( \frac{r_m}{r} \right)^{n(r)}
\! - {\frac{n(r)}{n(r) - m}} \left(\frac{r_m}{r} \right)^{m}
\right]
\end{equation}
\noindent
where $\epsilon$ is the potential depth, $r_m$ the position of the minimum and $n(r)$ is defined as follows~\cite{PBRCCV:PCCP08}:

\begin{equation}
\label{nILJ}
n(r) = \beta + 4 \left( \frac{r}{r_m} \right)^2.
\end{equation}

The corresponding parameters for the  potential using the ILJ analytical expression are given in \autoref{table1}.
A fine tuning has been carried out for both $\epsilon$ and $r_m$ by exploiting the comparison with the benchmark CCSD(T) interaction energies (not shown here)
until a quite good agreement between the {\it ab initio} estimations and their analytical representation is obtained.
Na$^+$ and He are bound through a pure non-covalent interaction, determined by the balance of size repulsion with induction and dispersion attraction, as also deduced by the obtained parameters in \autoref{table1}, which are in very good agreement  with those predicted by correlation formulas that exclusively exploit charge and polarizabilty of the interacting partners ($\epsilon$~=~45~meV and $r_m$~=~2.27~\AA, reported in Ref.~\cite{CPL:1991}).

The 3B non-covalent contribution to be added to the above described 2B counterpart is based on the dominant induced dipole-induced dipole interaction term, already employed in previous studies~\cite{liu_hyperspherical_2016,MCBGYYY:TCA07,RLAOPBHCGHBSG:PCCP18}, and here parameterized as:
\begin{eqnarray}
\label{eq:3}
V_{\rm 3B} & = & -\frac{\alpha^2}{2}
\left[
{\frac{3 r_j}{2}} g_3(r_i) g_5(r_{ij}) + {\frac{3 r_i}{2}} g_3(r_j) g_5(r_{ij})
\right. \nonumber \\
& - & {\frac{1}{2}} g_3(r_i) g_3(r_j) g_1(r_{ij})
-3 g_1(r_i) g_1(r_j) g_5(r_{ij}) \nonumber \\
& - & \left.
g_1(r_i) g_3(r_j) g_3(r_{ij}) - g_3(r_i) g_1(r_j) g_3(r_{ij})
\right]
\end{eqnarray}
\noindent
where $\alpha=1.42\, a_0^3$, $r_i$ and r$_j$ are \ce{Na+}--He distances, $r_{ij}$ is the He--He distance, and $g_n(r_i) = 1/r_i^n$.

\begin{table}[h]
\begin{center}
\small
\begin{tabular*}{0.5\textwidth}{@{\extracolsep{\fill}}lcccc}
  \toprule
     & $m$ & $\epsilon$ & $r_m$
      & $\beta$
      \\
    \hline
    Na$^{+}$--He & 4 &  43.0 meV & 2.31 \AA & 6.5    \\

    \bottomrule
  \end{tabular*}
\end{center}
  \caption{ILJ potentials parameters for the  Na$^{+}$--He interaction.  Note $m$ and $\beta$ are dimensionless.}\label{table1}
\end{table}
The evaporation energies of the Na$^+$He$_N$ complexes, defined
as $E_\text{evap}(N) = E_\text{bind}(N) - E_\text{bind}(N-1)$,
where $E_\text{bind}(N)$ is the total binding energy of Na$^+$He$_N$,
are then calculated using the above described PES
by means of path integral Monte Carlo (PIMC) calculations.
The same PIMC method~\cite{RGV:IRPC16}
has been successfully employed in previous investigations
of helium doped clusters~\cite{ZGLBHCGBZSPHB:JPCL23,BMPGHCHBS:M21,PMGSPHBOBGHCV:JCP19,RBHCGVPPHB:JCP17}
and details can be found elsewhere.
Here it suffices to say that the $E_\text{bind}(N)$ energies are obtained
using the thermodynamic estimator~\cite{RGV:IRPC16}
and that about 1200 quantum beads have been considered.
As usual, the PIMC simulations are carried out considering initial
geometries classically optimized by means of an evolutionary algorithm~\cite{I:CPC01}.

\begin{figure}
    \includegraphics[width = 8.6 cm]{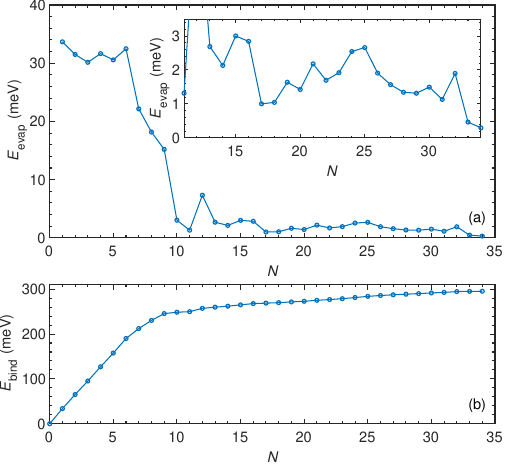}
    \caption{(a) The evaporation energy, $E_\text{evap}(N)$ of the \ce{Na+He_{$N$}} complexes calculated by PIMC as elaborated in \autoref{sec:EvaporationEnergy}. (b) The binding energy $E_\text{bind}(N)$ obtained using \autoref{eq:Sum}. \label{fig:EvaporationEnergy}}
\end{figure}

\Autoref{fig:EvaporationEnergy} displays $E_\text{evap}(N)$ for $N =1-34$.
The PIMC results show that the energies remain around 31~meV up to $N = 6$ where
a marked decrease is observed.
For Na$^+$He$_{12}$ a local maximum suggests the possible existence of a particularly stable structure, a feature which was also found for complexes of He atoms with other cations like Ca$^{2+}$~\cite{ZGLBHCGBZSPHB:JPCL23}.

\subsection{Attachment rate of He atoms}\label{sec:results:attachRate}
If we assume that the Na$^+$ ion binds to He atoms at a constant rate, $r$, and that the binding events are independent, then the probability, $P(t;n)$, that $n$ He atoms have bound to \ce{Na+} at time $t$ is given by:
\begin{equation}
    P_\text{Pois}(n;t)=\frac{(rt)^{n}\exp^{-rt}}{n!} \label{eq:poiss}
\end{equation}
i.e. a Poisson distribution. To determine $r$, we make a global fit of the Poisson distribution to the time-dependent ion yields. This strategy takes into account more of the actual data recorded compared to the method used in our first report~\cite{albrechtsenObservingPrimarySteps2023}, where the determination of $r$ was based solely on the peak positions of the $Y_n$($t$) traces.

We note the following two points prior to doing the fitting. Firstly, we include only the $Y_n$($t$) data points for $n \leq 5$. This is because the $Y_n$($t$) signals with $n~\geq~6$ are influenced by dissociation of \ce{Na+He_{$N$}} complexes with $N~>~6$ and such dissociation processes are not accounted for in the Poisson model. Dissociation of a \ce{Na+He_{$N$}} complex, i.e. \ce{Na+He_{$N$}}~$\rightarrow$~\ce{Na+He_{$N-1$}} + He, can occur after the complex leaves the He droplet if its internal energy, $E_\text{int}(N)$, exceeds $E_\text{evap}(N)$. For $N~\leq~6$, $E_\text{evap}(N)$ is sufficiently large, see \Autoref{fig:EvaporationEnergy}, that dissociation becomes negligible
as discussed in the Methods section of Ref.~\cite{albrechtsenObservingPrimarySteps2023}. By contrast,
the drop in $E_\text{evap}(N)$ for $N~>~6$, together with the fact that larger complexes can contain more internal energy, cause these complexes to undergo significant dissociation.

Secondly, when the Coulomb repulsion begins, it takes a little time before the \ce{Na+He_{$N$}} ion has moved far enough away from the droplet to stop interacting with it, and in this time the \ce{Na+} ion can bind more He atoms. As a first approximation, we consider this lift-off time equal for all Na$^+$He$_N$ ions and denote it $t_\text{lift}$. Taking $t_\text{lift}$ into account and, furthermore, multiplying the Poissonian probability density function by an amplitude $A$ to match the experimental ion yields, gives
\begin{equation}
    f(t; n) = A\frac{(r(t-t_\text{lift}))^{n}\exp^{-r   (t-t_\text{lift})}}{n!}.\label{eq:poissFitFun}
\end{equation}
The parameters $A$, $t_\text{lift}$ and $r$ were optimized by a global least-squares fit of \autoref{eq:poissFitFun} to the first 10 ps of the $Y_n$($t$) curves for $n \leq 5$. The resulting fit, shown by the full blue lines in \autoref{fig:yields}(a0)-(a5), is obtained for an attachment rate $r$ of $1.84\pm0.09$ He atoms pr. ps. The fit describes the data very well. We note that this value differs slightly from the rate of 2.0 He atoms pr. ps reported in Ref.~\cite*{albrechtsenObservingPrimarySteps2023}. As mentioned above, the previous result is based only on the center position of the peaks, while the new result is based on a fit to the full yield curve, and we believe this new method to be more accurate.

We extrapolated the resulting fit beyond the data range used for the fit itself. \Autoref{fig:yields}(a6)-(a9) show that for $n~=~6-9$, the curves from the Poissonian fit, now shown by the dashed blue lines, still resemble the experimental results but underestimate $Y_n$($t$). The reason is the dissociation processes mentioned before. For $N~=~6-9$, $E_\text{evap}(N)$ drops strongly and, therefore, the dissociation probability increases as $N$ increases. Thus, $Y_n$($t$) receives a contribution from dissociation of larger complexes which exceeds the loss due to dissociation of the \ce{Na+He_{$N=n$}} ions. For the even larger complexes with $n~=~10-14$, it is seen that the fit overestimates $Y_n$($t$). This is due to the fact that for $N~=~10-14$, $E_\text{evap}(N)$ is so low, see \autoref{fig:EvaporationEnergy}, that even a small amount of internal energy in a \ce{Na+He_{$N$}} complex leads to evaporation of several He atoms.  Notably, the large decrease in $E_\text{evap}(N)$ from $N$~=~9 to $N$~=~10 causes a pile up of signal in $Y_9$($t$). For more details, see the Methods section of Ref.~\cite{albrechtsenObservingPrimarySteps2023}.

We note that for $n \geq 6$, the curves from the Poissonian fit underestimate the $Y_n$($t$) curves at early times (0-3 ps). One possible explanation is many-body binding processes where several He atoms attach to the ion simultaneously. Such processes can account for why e.g. \ce{Na+He_9} ions are formed in non-negligible amounts already at $t$~=~2.0~ps, a time when the formation probability would be essentially zero if solvation was the sole result of sequential binding events with the measured rate. In Ref.~\cite{albrechtsenObservingPrimarySteps2023}, time-dependent DFT simulations showed that already after 1-2 ps, the \ce{Na+} ion is uniformly surrounded by liquid helium on its droplet-facing half. This close contact between the ion and the He atoms would appear favorable for many-body interactions.

\begin{figure*}
    \includegraphics[width = 17.8 cm]{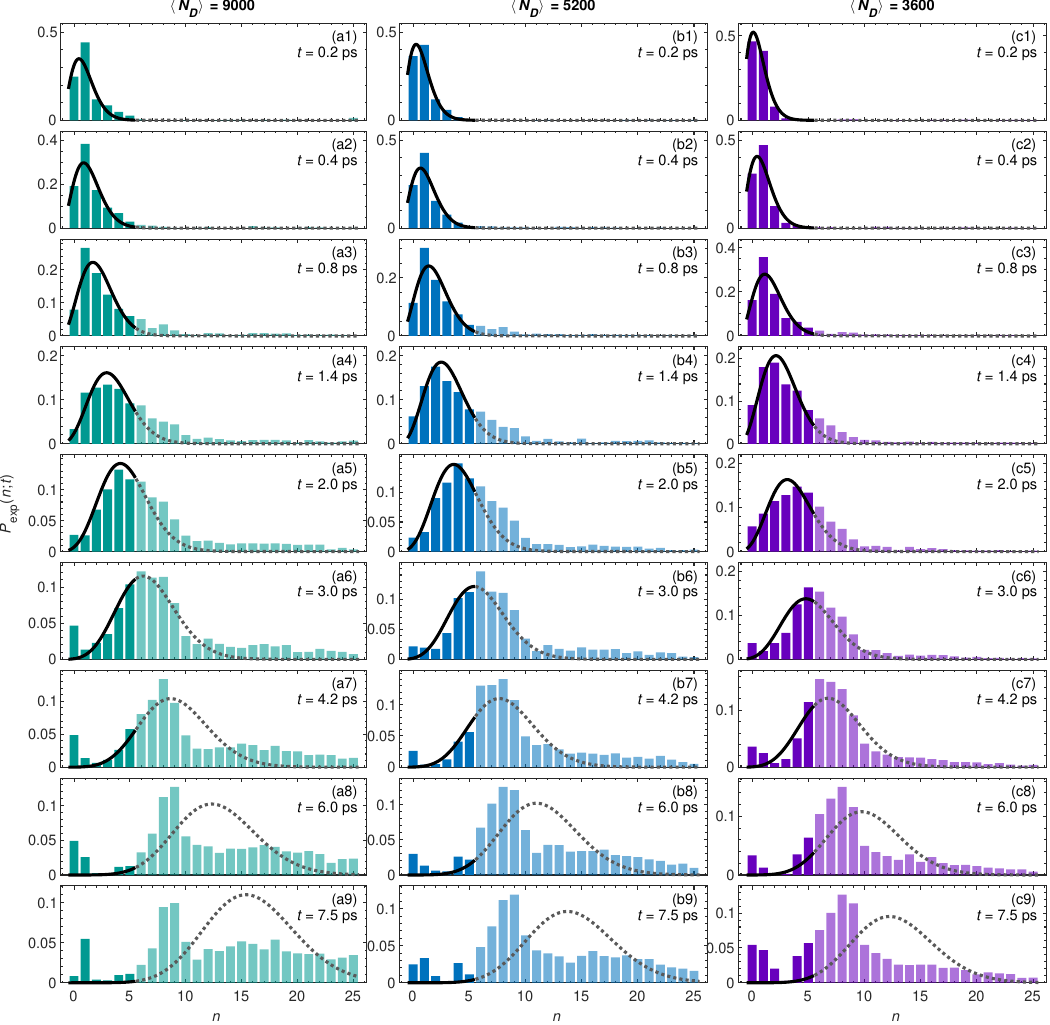}
    \caption{Normalized number distributions, in a histogram representation, of the \ce{Na+He_{$n$}} complexes at nine selected times and for three different droplet size distributions ($\avg{N_D}$ is given above each column).  Full black lines: Results from the fit to the Poisson distribution, see text and \autoref{fig:yields}. The dark blue bars represent the experimental data included in the fit. The dotted grey line is the extrapolation of the fit beyond the fitted range. \label{fig:nDist_comparison}}
\end{figure*}

The $Y_n$($t$) signals shown in \autoref{fig:yields}, enable us to directly determine the distribution, $P_\text{exp}(n;t)$, of the \ce{Na+He_{$n$}} ion signals at a given time. \Autoref{fig:nDist_comparison}(b1)-(b9) depicts the results for nine selected times. At $t$~=~0.2~ps, the distribution is dominated by the \ce{Na+} and \ce{Na+He} signals. This makes sense since the newly formed \ce{Na+} ion has had only little time to effectively bind any He atoms. As time elapses, the number distributions grow broader and their weight shifts to larger $n$, a behavior illustrating the statistical nature of the gradual \ce{Na+} ion solvation with the He atoms.

\Autoref{fig:nDist_comparison}(b1)-(b9) also show the outcome of the fit applied to the number distributions, i.e. \autoref{eq:poissFitFun} with the parameters $A$, $t_\text{lift}$ and $r$ found before.  As for the time-dependent ion signals, \autoref{fig:yields}, the fitted results (full black curves) match the experimental data well in the $n$~=~0-5 range, which the global fit is based on, while deviations appear beyond this range where the fit is extrapolated (black dotted curves). Notably, at $t$~=~6.0~ps and $t$~=~7.5~ps, the experimental number distributions peak at significantly lower $n$-values than the fitted curves. In line with the discussion of the $Y_n$($t$) curves, we ascribe this as being due to dissociation of the larger \ce{Na+He_{$N$}} complexes.

\begin{table}
    \centering
    \begin{tabular}{ccc}
    \toprule
    $\avg{N_D}$ & $r$ (He/ps) & $t_\text{lift}$ (ps) \\ \midrule
    3600      & $1.65\pm0.09$        & $-0.19\pm0.06$            \\
    5200      & $1.84\pm0.09$        & $-0.24\pm0.06$            \\
    9000      & $2.04\pm0.13$        & $-0.28\pm0.07$            \\
    \bottomrule
    \end{tabular}
    \caption{The rate and the lift-off time determined from the fit to the data sets for the three different average droplet sizes explored. All uncertainties were extracted from the curve fits and are given as two standard deviations.}
    \label{tab:poissonResults}
\end{table}

As previously mentioned, the experiment was also carried out for droplets with an average size, $\avg{N_D}$, of 3600 and 9000. The time-dependent ion yields (see supplementary material) were analyzed in the same way as for the $\avg{N_D}~=~5200$ data set discussed above. The resulting number distributions $P_\text{exp}(n;t)$ and fits are shown in \autoref{fig:nDist_comparison}(a1-a9) and (c1-c9). It can be seen that the Poisson fit match the experimental data well for $n~<~6$ and for times up to about 4 ps, quite similar to the $\avg{N_D}~=~5200$ case. Comparing the results for the three values of $\avg{N_D}$ allows us to investigate the effect of the droplet size on the solvation dynamics. The results in \autoref{tab:poissonResults} show that larger droplets lead to larger (more negative) time shifts. This behavior is consistent with the expected slower ejection time of \ce{Na+He_$N$} from the droplet as the Coulomb repulsion is weaker for larger droplets due to the increased distance between \ce{Xe+} and \ce{Na+}.  The fit also shows that the helium attachment rate increases when the droplet size increases. This could be because the increase of $t_\text{lift}$ with droplet size leaves \ce{Na+He_$N$} more time to pick up additional He atoms and to dissipate energy for larger complexes. Another explanation for this observation is that larger droplets may be able to dissipate the solvation energy faster since the density of states in the droplet increases with size~\cite{brinkDensityStatesEvaporation1990}. Thirdly, the softer Coulomb explosion for the larger droplets will add less vibrational energy to the ejected \ce{Na+He_{$N$}} ions. As such, the ions will have a smaller probability for evaporating He atoms, which in turn leads to detection of larger \ce{Na+He_{$n$}} complexes at earlier times. Currently, we are not able to distinguish between these three mechanisms.

At this point, we would like to mention results that are somewhat related to the findings presented here. In Ref.~\cite{schulz_formation_2001}, Schulz, Claas and Stienkemeier electronically excited K atoms ($4s \rightarrow 4p$), residing at the surface of He droplets with a fs pump pulse and measured the time-resolved formation of a K*He exciplex by recording fluorescence induced by a fs probe pulse, where K* denotes K(4p). They found that the abundance of the exciplex reached a maximum after 180 fs and then gradually decays in another 200 fs. Although time-resolved data were shown only for K*He, the authors reported observation of larger exciplexes K*He$_n$. It would be interesting to repeat this experiment and see if it is possible to observe solvation of K* (or other excited alkali atoms) by several He atoms, similar to what we measure for the \ce{Na+} ion in the current work.

\subsection{Time-dependent energy dissipation}\label{sec:results:energyDisp}

As discussed in \autoref{sec:results:attachRate}, quantitative characterization of the time-dependent attachment of He atoms to the \ce{Na+} ion is less accurate for \ce{Na+He_{$n$}} complexes with $n > 5$ because they dissociate after leaving the droplet surface. In this section, we show how these dissociation processes make it possible to determine the energy dissipated from the local region around the \ce{Na+} ion as a function of time. During the formation of a Na$^+$He$_N$ complex, the binding energy of the Na$^+$--He bonds is dissipated into the droplet and via He atoms ejected from the surface~\cite{albrechtsenObservingPrimarySteps2023}. As detailed below, the size of the \ce{Na+He_{$n$}} ion detected provides information about the time-dependence of the energy dissipated.

Energy conservation dictates that at any time:

\begin{align}
    E_\text{init}(\ce{Na+}) = E_\text{disp} + E_\text{bind}(N) + E_\text{int}(N)
\end{align}

and thus:
\begin{align}
    E_\text{disp} = -E_\text{bind}(N) - E_\text{int}(N) + E_\text{init}(\ce{Na+}). \label{eq:EDisp:EDispInit}
\end{align}
where $E_\text{init}$(\ce{Na+}) is the energy of the \ce{Na+} ion created at $t=0$, $E_\text{disp}$ the energy dissipated, and $E_\text{bind}(N)$ and $E_\text{int}(N)$ the binding energy and internal energy, respectively, of the \ce{Na+He_{$N$}} complex. To ease the notation, we omit the explicit indication of time dependence, ($t$), on the $E_\text{disp}$ and $E_\text{int}(N)$ terms.

As mentioned previously, a \ce{Na+He_{$N$}} complex can lose one or more He atoms, by evaporation, if $E_\text{int}(N) > E_\text{evap}(N)$. To deduce $E_\text{disp}$ from the size of the \ce{Na+He_{$n$}} ions detected, we make three assumptions about the evaporation events:

I) The evaporated He atoms have a negligible amount of kinetic energy. Thus, the sum of the binding energy and the internal energy of the complex is conserved:
\begin{align}
    E_\text{bind}(N) + E_\text{int}(N) \nonumber \\
    = E_\text{bind}(n) + E_\text{int}(n) \label{eq:EDisp:EtotalConserved}
\end{align}
II) Evaporation continues until the remaining internal energy is too low to allow another evaporation event:
\begin{align}
    0 \leq E_\text{int}(n) < E_\text{evap}(n) \label{eq:EDisp:EintFinal}
\end{align}
III)
The evaporation of He atoms happens much faster than the travel time of the ions between the repeller and extractor electrodes of the VMI spectrometer. This ensures that the \ce{Na+He_{$n$}} ion measured at the detector is the one where the initial \ce{Na+He_{$N$}} complex has evaporated off as many He atoms as it can.

To justify the three assumptions, the free \ce{Na+He_{$N$}} complex was simulated using a molecular dynamics (MD) simulation for two initial complex sizes and three initial internal energies. The simulations were carried out with Tinker \cite{rackersTinkerSoftwareTools2018} using the AMOEBA-09\cite{ren_polarizable_2011} force field. The force field for the Na$^{+}$--He interaction was fitted to the corresponding potential from Ref. \cite{koutselosInteractionUniversalityScaling1990},\footnote{Thus, the Na$^{+}$--He interaction used in the MD simulation differs slightly from the one used for the PIMC calculation in Sec. IV B}.

First, a \ce{Na+He_{$N$}} complex is generated and its energy is minimised, using a built-in algorithm in Tinker (Limited-Memory Broyden-Fletcher-Goldfarb-Shanno), until it is completely vibrationally relaxed. Now, a specific amount of initial internal (vibrational) energy, $E_\text{int}^{(0)}$, is instantly added to the complex by giving each helium atom a velocity. The allocation of the velocities is done in a random manner requiring, however, that the sum of the kinetic energies of all the He atoms equals $E_\text{int}^{(0)}$ and avoiding that a single He atoms gets all the kinetic energy. The latter is obtained by controlling the maximal and minimal energy fraction each helium atom can gain.

Second, the MD simulation is run for \SI{1}{ns} in time steps of \SI{1}{fs}. The internal energy of the complex, $E_\text{int}(t)$, is determined every 100 fs for the first 10 ps and every 10 ps thereafter. If enough kinetic energy localises in a single helium atom, we observe that this atom leaves the complex. In practice, the He atom is considered to have left, i.e. to be no longer bound, if it has moved more than 10 {\AA} away from the Na$^{+}$ ion. Such an event lowers the internal energy of the remaining complex by the evaporation energy and by the final kinetic energy of the escaped He atom, the latter denoted $E_\text{kin}^{\text{He}}$. The simulations are run for three values of $E_\text{int}^{(0)}$: \SI{37}{meV}, \SI{124}{meV} and \SI{248}{meV}, and for two different sizes of the initial complex: Na$^{+}$He$_6$ and Na$^{+}$He$_{14}$. Each simulation is repeated 1000 times, i.e. for 1000 different ways of distributing $E_\text{int}^{(0)}$ among the $n$ He atoms. For each of the two complexes starting with $E_\text{int}^{(0)}$, the average internal energy, $\langle E_\text{int}\rangle(t)$ is found by averaging over the 1000 simulations at each time step.

\begin{figure}
    \includegraphics[width = 8.6 cm]{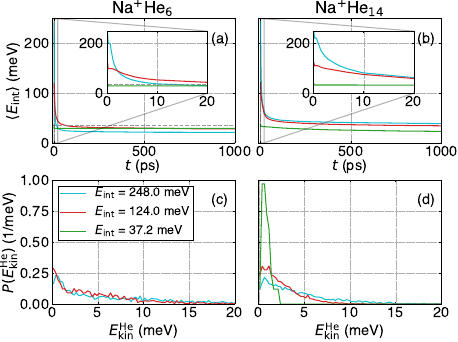}
    \caption{Results of the molecular dynamics simulation for complexes \ce{Na+He_6}, (a) and (c), and \ce{Na+He_14}, (b) and (d) with different initial internal energies $E_\text{int}^{(0)}$ . (a)-(b): The time dependence of the average internal energy, $\langle E_\text{int}\rangle$, of the remaining complex. In (a), the dashed line $\avg{E_\text{int}} = \SI{35}{\meV}$ marks the approximately constant $E_\text{evap}(N)$ for $N \leq 6$. (c)-(d): The corresponding distributions of the kinetic energies of the evaporated helium atoms. \label{fig:MDSim}}
\end{figure}

The results, displayed in \autoref{fig:MDSim}(a)-(b), show that $\avg{E_\text{int}}(t)$ quickly falls to an asymptotic value in a matter of few tens of picoseconds, which reflects the fact that the evaporation process is finished on this timescale. The size of the measured ion complexes is determined by TOF mass spectrometry in the VMI spectrometer, as described in \autoref{sec:setup}. A Simion~\cite{dahl_simion_2000} simulation of dissociating complexes in the VMI spectrometer was made, and it was found that the measured complex size is always the final size after dissociation, even if the dissociation happens after \SI{100}{\ns}, much longer than what is found from the MD simulation. Therefore, the Na$^{+}$He$_n$ ion measured is the complex after it has dissociated as many times as its initial internal energy allowed it to do. This justifies assumption III.

Furthermore, the asymptotic values of $\avg{E_\text{int}}(t)$ shown in \autoref{fig:MDSim}(a) are just below the constant evaporation energy of \SI{35}{\meV} found for $N \leq 6$ with the numerical method used in the MD simulation~\footnote{The \ce{Na+}--He potential used in the MD simulations differs somewhat from that used in the PIMC calculation and, therefore, the evaporation energies determined in the MD simulations are a little different from the PIMC results. This difference is, however, not important for the conclusions drawn from the MD simulation}. For the initial \ce{Na+He_{14}} complex, $\avg{E_\text{int}}(t)$,  shown in \autoref{fig:MDSim}(b), is harder to interpret as the different possible final complexes have much different evaporation energies. The asymptotic $\avg{E_\text{int}}(t)$ values do, however, seem consistent with the evaporation energies of the final complexes expected based on the initial internal energies. These results are consistent with assumption II, that the evaporation process continues until the remaining internal energy is below the evaporation energy.

\Autoref{fig:MDSim}(c)-(d), displaying the distribution of $E_\text{kin}^\text{He}$, show that the majority of the evaporated helium atoms carry off only a few \si{\meV} of kinetic energy, which keeps the total energy of the complex approximately constant. This justifies assumption I.

\begin{figure*}
    \includegraphics[width = 17.8 cm]{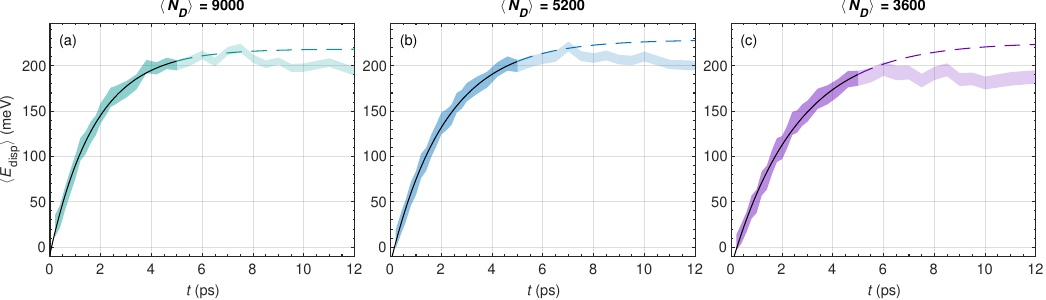}
    \caption{The upper and lower bounds of the mean energy dissipated to the droplet as a function of time for three different average droplet sizes. The darker colored region $t \leq \SI{5}{\ps}$ signifies the range in which $\avg{E_\text{disp}}$ is expected to be well estimated, see text. In each panel, the solid black line is an exponential fit using \autoref{eq:EDisp:fitFunction}, while the dashed colored line is an extrapolation of this fit beyond the fitted range. \label{fig:EDispComparison}}
\end{figure*}

To proceed, equations \eqref{eq:EDisp:EDispInit}, \eqref{eq:EDisp:EtotalConserved} and \eqref{eq:EDisp:EintFinal} are combined:
\begin{align}
    -E_\text{bind}(n-1) + E_\text{init}(\ce{Na+}) \leq E_\text{disp} \nonumber \\
    < -E_\text{bind}(n)+ E_\text{init}(\ce{Na+}), \label{eq:EDisp:EDispRange}
\end{align}
where we used that the binding energy can be written as the sum over the evaporation energies
\begin{align}
    E_\text{bind}(n) = \sum_{i=1}^{n} -E_\text{evap}(i). \label{eq:Sum}
\end{align}
\Autoref{eq:EDisp:EDispRange} shows that when a \ce{Na+He_$n$} ion is detected, an energy between $-E_\text{bind}(n-1) + E_\text{init}(\ce{Na+})$ and $-E_\text{bind}(n)+ E_\text{init}(\ce{Na+})$ was removed from the local region of the \ce{Na+} ion to form the parent \ce{Na+He_{$N$}} complex while \ce{Na+} was still in contact with the droplet.

As illustrated in \autoref{fig:nDist_comparison}, we measure a distribution of \ce{Na+He_{$n$}} ions at any time. This distribution, termed $P_\text{exp}(n;t)$, must be accounted for and, thus, we determine the mean energy dissipated, $\langle E_\text{disp} \rangle$, at time $t$ as the weighted sum of $E_\text{disp}$ over $P_\text{exp}(n;t)$. Using \autoref{eq:EDisp:EDispRange} we obtain:
\begin{align}
    \sum_{i=1}^{25} P_\text{exp}(i;t)(-E_\text{bind}(i-1)+ E_\text{init}(\ce{Na+})) \leq \avg{E_\text{disp}} \nonumber \\
    < \sum_{i=1}^{25} P_\text{exp}(i;t)(-E_\text{bind}(i)+ E_\text{init}(\ce{Na+})). \label{eq:EDisp:Average}
\end{align}
with the sum ending at $n = 25$ because, as mentioned, we cannot measure the yield of the \ce{Na+He_{$n$}} ions for $n > 25$.

\Autoref{fig:EDispComparison} displays the results for $\avg{N_D} =$ 3600, 5200, and 9000 with the lower (upper) edge of the bands indicating the lower (upper) limit for $\langle E_\text{disp} \rangle$ as given in \autoref{eq:EDisp:Average}. The shapes of the $\langle E_\text{disp} \rangle$ curves are similar for the three different droplets sizes with a steep initial rise followed by a gradual flattening, i.e. the rate of the energy dissipation decreases as a function of time. This behavior brings to mind Newton's law of cooling where the rate of heat transfer of a hot (macroscopic) body to its surroundings is proportional to the internal energy of the body. As a result, the internal energy decreases exponentially in time and stabilizes at a value determined by the temperature of the environment.

To test if $\langle E_\text{disp} \rangle$(t) follows Newton's law of cooling, we fitted the experimental data points shown in \autoref{fig:EDispComparison} to the function:
\begin{align}
    E_\text{disp}^\text{Newt}(t) = E_\text{disp}(\infty)\left( 1 - \exp\left( -\frac{(t-\Delta t_\text{disp})}{\tau_\text{disp}} \right)\right), \label{eq:EDisp:fitFunction}
\end{align}
where $E_\text{disp}(\infty)$ is the asymptotically dissipated energy, $\tau_\text{disp}$ is the time constant for the energy dissipation, and $\Delta t_\text{disp}$ is a time offset analogous to $t_\text{lift}$ described earlier. Since we are not able to measure the yield of the \ce{Na+He_{$n$}} ions for $n > 25$, we only consider $\langle E_\text{disp} \rangle$ to be accurate in the time range where the $Y_{25}$ signal is negligible. With reference to \autoref{fig:nDist_comparison}, this is the case when $t\leq\SI{5}{\ps}$ and thus the fit is only done for $0 < t \leq 5 \text{ps}$.

The full lines in \autoref{fig:EDispComparison} represent the best fits to the data points in the interval $0 - 5~\text{ps}$ and the dashed lines the fit beyond the fitted range, i.e. for $ t > 5~\text{ps}$. Firstly, we note that \autoref{eq:EDisp:fitFunction} describes $\avg{E_\text{disp}}$($t$) very well within the fitted range for all three droplets sizes. This indicates that the dynamics of the energy dissipation in the \ce{Na+} solvation process is similar to that expressed by Newton's law of cooling. Secondly, $E_\text{disp}(\infty)$, given in \autoref{tab:expfitResults}, is approximately the same for the three droplet sizes. This makes sense because we interpret $E_\text{disp}(\infty)$ as the total solvation energy of the \ce{Na+} ion in the helium droplet and that energy is expected to be essentially droplet size independent for the sizes considered here. However, the absolute value of $\sim\SI{224}{\meV}$ is only $\sim 58 \%$ of the total solvation energy, \SI{384}{\meV}, calculated by density functional theory~\cite{garcia-alfonso_time-resolved_2024}. We believe the major reason for the difference is that the Coulomb repulsion adds internal energy to the \ce{Na+He_{$N$}} complex during its ejection from the droplet. An increased internal energy leads to more evaporation of the ejected \ce{Na+He_{$N$}} complexes, and thus the \ce{Na+He_{$n$}} ions detected will be smaller on average. As such, our measurement tends to underestimate the true dissipated energy. The effect will be more pronounced for the larger complexes, and therefore it is expected to influence the measurement of the dissipated energies more at later times.

Thirdly, the results in \autoref{tab:expfitResults} show that the time constant for the energy dissipation expressed by $\tau_\text{disp}$, depends on the droplet size, with larger droplets displaying a faster measured energy dissipation. This observation may to some degree be related to the gradually increasing underestimate of the energy dissipated as the droplet sizes increases and, as mentioned above, the increase of $t_\text{lift}$ with droplet size and thus more time to dissipate energy. However, the observation also points to that larger droplets are able to absorb energy more quickly than smaller droplets. This could be explained by that fact that larger droplets have a higher density of states and surface modes to accept the energy~\cite{brinkDensityStatesEvaporation1990} from the solvation process.

Finally, the time offset $\Delta t_\text{disp}$ shows a clear trend, with larger droplets leading to smaller time shifts. In contrast to the time offset $t_\text{lift}$ from the Poisson fit to $Y_n(t)$,
$\Delta t_\text{disp}$ is positive for all droplet sizes. This could be due to the Coulomb repulsion inducing internal energy in the \ce{Na+He_{$N$}} complexes. This induced energy is not part of the modelling used,
but would have to be dissipated in addition to binding energy of the He atoms, before we would measure a dissipated energy $E_\text{disp}$, leading to a positive time offset $\Delta t_\text{disp}$. Smaller
droplets lead to a stronger Coulomb repulsion and thereby to a larger induced internal energy, explaining the trend in $\Delta t_\text{disp}$.

\begin{table}
    \centering
    \begin{tabular}{cccc}
    \toprule
    $\avg{N_D}$ & $\tau_\text{disp}$ (ps) & $\Delta t_\text{disp}$ (ps) & $E_\text{disp}(\infty)$ (\si{\meV}) \\ \midrule
    3600      & $2.6\pm0.4$        & $0.23\pm0.06$       & $226\pm16$ \\
    5200      & $2.1\pm0.2$        & $0.15\pm0.04$       & $229\pm8$ \\
    9000      & $1.8\pm0.2$        & $0.06\pm0.06$       & $219\pm8$ \\
    \bottomrule
    \end{tabular}
    \caption{Comparison of the results obtained when fitting the exponential model see \autoref{eq:EDisp:fitFunction} and \autoref{fig:EDispComparison} to $\langle E_\text{disp} \rangle (t)$. All uncertainties were extracted from the curve fits and are given as two standard deviations.}
    \label{tab:expfitResults}
\end{table}

\section{Conclusion}\label{sec:conclusion-outlook}

We explored the primary steps of solvation of a single sodium ion in a helium droplet by measuring the binding of the first few solvent atoms to \ce{Na+} and the energy dissipated from its local surroundings, as a function of time. A Poisson analysis of the time-dependent \ce{Na+He_{$n$}} ion yields showed that the binding of the first five He atoms to the \ce{Na+} ion is well-described by a Poissonian process with a binding rate of $1.65\pm0.09$ ps$^{-1}$, $1.84\pm0.09$ ps$^{-1}$, and $2.04\pm0.13$ ps$^{-1}$, for droplets containing an average number of 3600, 5200 and 9000 He atoms, respectively. The difference in the rates could be due to the solvation process occurring faster in larger droplets but also because larger droplets lead to a more gentle Coulomb repulsion of the \ce{Na+He_{$N$}} complexes from the surface. Based on three assumptions, justified by MD simulations, about the evaporation dynamics of He atoms from hot \ce{Na+He_{$N$}} complexes departing from the droplet surface, we determined the average energy dissipated from the local region of the \ce{Na+} ion at each time step recorded. We found that for the first 5 ps, the time dependence of $E_\text{disp}$ follows Newton's law of cooling. The energy dissipation occurs faster as the droplet size increases, which may partly be due to larger droplets having a higher density of acceptor modes.

\section*{DATA AVAILABILITY}
The source data presented in this work and the analysis scripts used to treat them can be obtained from the authors on reasonable request.

\section*{Supplementary Material}
Figure showing the time-dependent yields of the different \ce{Na+He_{$n$}} ions, $Y_n$($t$), for droplets with $\avg{N_D} = $ 9000 and 5200.

\begin{acknowledgments}
We thank Jan Th{\o}gersen for expert help on keeping the laser system in optimal condition. We are grateful to Laurits Lønskov Sørensen and Simon Fischer-Nielsen for their great help in writing the optimized program for clustering and centroiding the TPX3Cam data. H.S. acknowledges support from Villum Fonden through a Villum Investigator Grant No. 25886.
\end{acknowledgments}

\vspace{2 cm}

\newpage


%


\clearpage
\widetext
\begin{center}
\textbf{\large Supplemental Materials: Femtosecond-and-atom-resolved solvation dynamics of a Na$^+$ ion in a helium nanodroplet}
\end{center}
\setcounter{equation}{0}
\setcounter{figure}{0}
\setcounter{table}{0}
\setcounter{page}{1}
\makeatletter
\renewcommand{\theequation}{S\arabic{equation}}
\renewcommand{\thefigure}{S\arabic{figure}}
\renewcommand{\bibnumfmt}[1]{[S#1]}
\renewcommand{\citenumfont}[1]{S#1}
\enlargethispage*{3cm}

\begin{figure*}[b]
  \includegraphics[width = 17.8 cm]{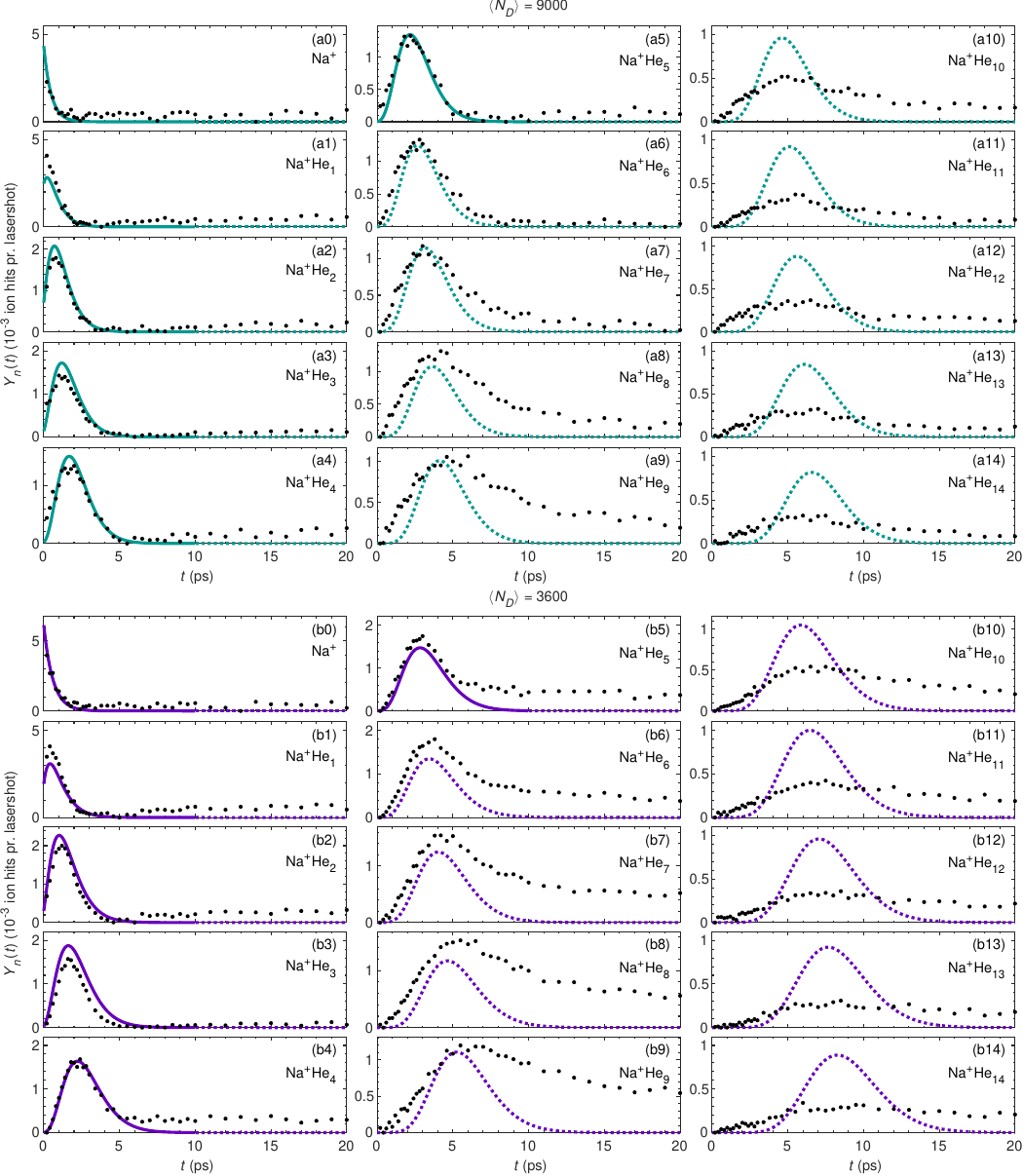}
  \caption{Black dots: Yields of different \ce{Na+He_{$n$}} ions, $Y_n$($t$),  measured as a function of pump-probe delay, $t$, using droplets doped with both a Na and a Xe atom and with $\avg{N_D} = $ 9000 (a0-a14) and 3600 (b0-b14). Full lines: Results from the fit to the Poisson model. The dotted lines show the extrapolation of this fit beyond the fitted range, see text for further details. \label{fig:SI}}
\end{figure*}

\end{document}